\documentclass[twocolumn]{aastex631}

\shortauthors{Echibur\'u-Trujillo \& Dexter}
\usepackage{amsmath}

\begin{document}

\title{Revealing the Accretion Flow in M87*: Insights from Faraday Rotation}

\author[0000-0001-8436-1847]{Constanza Echibur\'u-Trujillo}
\affiliation{Department of Astrophysical and Planetary Sciences, JILA, Duane Physics Bldg., 2000 Colorado Ave., University of Colorado, Boulder, CO 80309, USA}
\correspondingauthor{C. Echibur\'u-Trujillo}
\email{constanza.echiburu@colorado.edu}

\author[0000-0003-3903-0373]{Jason Dexter}
\affiliation{Department of Astrophysical and Planetary Sciences, JILA, Duane Physics Bldg., 2000 Colorado Ave., University of Colorado, Boulder, CO 80309, USA}

\begin{abstract} 
The Faraday rotation measure (RM) is a commonly used tool to trace electron number density and magnetic fields in hot accretion flows, particularly in low-luminosity accreting supermassive black holes. We focus on the nuclear region of M87, which was observed at 230 GHz (1.3 mm) by the Event Horizon Telescope in 2019. It remains unclear whether this emission originates from the accretion flow, the jet base, or both. To probe the presence of an accretion flow, we explore the scenario where the linearly polarized emission from the counter-jet, visible at 43 GHz (7 mm), is Faraday-rotated by the accretion flow. We calculate theoretical predictions for counter-jet polarization using analytical and numerical models. In all cases, we find a Faraday-thick flow at 43 GHz (7 mm), with $\mathrm{RM} \sim 10^6$ rad m$^{-2}$, and a polarization angle that follows a linear relationship with wavelength squared, consistent with external Faraday rotation. The more realistic model, which includes turbulence and magnetic field fluctuations, predicts that the polarization pattern should be time-dependent, and that the counter-jet emission is depolarized due to Faraday depth fluctuations across the accretion flow. Despite the Faraday thick regime and strong depolarization, the linear relationship persists, enabling us to constrain the flow´s physical properties. Comparing the counter-jet and forward-jet linear polarization states should enable detection of M87's accretion flow and provide lower limits on electron density, magnetic field strength, and mass accretion rate.

\end{abstract}


\section{Introduction} \label{sec:intro}

Most large galaxies containing supermassive black holes (SMBHs) at their cores are significantly underluminous compared to their bright counterparts, Active Galactic Nuclei. This suggests that these objects may be powered by a different mode of accretion, where the mass accretion rate $\dot{M}$ is low \citep[$\lesssim 0.1 \dot{M}_{\mathrm{Edd}}$, e.g.,][where $\dot{M}_{\mathrm{Edd}}$ is the Eddington accretion rate]{2009ApJ...699..626H} and decreases from the outer accretion disk towards the SMBH \citep[see][for a review]{2012ApJ...761..129Y}, implying that gas is lost along the way. In this scenario, the low gas density prevents it from cooling efficiently, leading to a radiatively inefficient accretion flow \citep[RIAF,][]{1994ApJ...428L..13N,1995ApJ...452..710N,1995ApJ...438L..37A}. Examples of such objects include Sgr A* in our Galaxy and M87* in the elliptical galaxy M87.

General Relativistic Magnetohydrodynamic Simulations (GRMHD) can realize RIAF accretion flow solutions computationally while including self-consistent evolution of magnetic fields and magnetorotational turbulence \citep{mri} from an initial condition. Detailed radiative models provide a good match to many event-horizon-scale observations of Sgr A* and M87*.

The giant elliptical galaxy M87 is located at a distance of 16.7 Mpc \citep{2009ApJ...694..556B,2010A&A...524A..71B} and, at its center, it hosts a SMBH with a mass of $(6.5 \pm 0.7) \times 10^9 \, M_{\odot}$ \citep{2019ApJ...875L...6E}, defining a scale of 1 mas $\approx 131 \, r_{\mathrm{S}}$. On larger scales, M87 exhibits a kiloparsec-scale relativistic jet almost directed towards the Earth \citep[$\theta = 17^{\circ}$,][hereafter forward-jet]{2018ApJ...855..128W} which has been observed at different electromagnetic wavelengths, and thus, scales \citep[e.g.,][]{2013ApJ...775...70H,2015MNRAS.451..588R,2016ApJ...817..131H,2016A&A...595A..54M,2019ApJ...871..257P}. In contrast, the counter-jet moving in the opposite direction is much weaker and observable only at subparsec scales \citep{2018ApJ...855..128W,2021ApJ...922..180P}. The Event Horizon Telescope (EHT) observations at 43 GHz (7 mm) have resolved part of the counter-jet structure. Presumably this jet is powered by mass accretion, but even in the highest resolution images at 230 GHz (1.3 mm), it remains unclear whether the observed emission originates from inflow or outflow of material (i.e., the accretion flow or the jet base, or both).

Faraday rotation provides another means to detect the presence of an accretion flow. As polarized light passes through a magnetized medium, its plane of polarization rotates by an amount that depends on the physical properties of that medium and the emission wavelength. The Faraday rotation effect has an observable signature known as the Faraday rotation measure (RM), defined by:

\begin{equation}
    \label{eq:rmobs1}
    \mathrm{RM} = \frac{\mathrm{EVPA}_2-\mathrm{EVPA}_1}{\lambda^2_2-\lambda^2_1},
\end{equation}

\noindent where the subscripts 1 and 2 represent two emission wavelengths ($\lambda$) and two measurements of the electric vector position angle (EVPA). When the magnetized medium (or Faraday ``screen'') is located between the observer and the source of polarized light, such that rotation occurs outside the emission source (i.e., \textit{external} Faraday rotation), the RM can be written as:

\begin{equation}
    \label{eq:rm}
    \Bigg( \frac{\mathrm{RM}}{\mathrm{rad \, \,  m}^{-2}} \Bigg) = 8.1 \times 10^5 \int_{\mathrm{src}}^{\mathrm{obs}} \Big( \frac{n_e}{\mathrm{cm}^{-3}} \Big) \Big( \frac{B_{||}}{\mathrm{G}} \Big) \Big( \frac{dl}{\mathrm{pc}} \Big),
\end{equation}

\noindent where $n_e$ is the electron number density, $B_{||}$ is the magnetic field component along the line of sight, and $dl$ is the path length between the source and the observer. From Equations \ref{eq:rmobs1} and \ref{eq:rm}, it follows that if Faraday rotation is external and EVPA measurements are made at closely separated wavelengths, then one can learn about the properties of the magnetized plasma, i.e., $n_e$ and $B_{||}$. The effect is maximized at low temperature and high density, exactly the conditions expected for a relatively cold accretion flow that is sub-dominant in synchrotron emission.

Theoretically, the EVPA rotation is quantified by the Faraday depth $\tau_{\rho_\mathrm{V}}$, where $\rho_\mathrm{V}$ is the Faraday rotation coefficient. When $\tau_{\rho_{\mathrm{V}}}\lesssim 1$ (or the EVPA rotates by $<180^{\circ}$), the plasma is considered Faraday thin and the EVPA can be traced back to its original position. In this regime, $\tau_{\rho_{\mathrm{V}}} = 2 \, \mathrm{RM} \lambda^2$ \citep[see][ Appendix C]{2016MNRAS.462..115D}. Thus, the RM provides information about $\tau_{\rho_\mathrm{V}}$, and when combined with Equations \ref{eq:rmobs1} and \ref{eq:rm}, it can be used to probe the plasma properties along the line of sight. However, when $\tau_{\rho_{\mathrm{V}}} \sim 1$ or larger, the plasma is Faraday thick, and the EVPA has rotated multiple times, making the polarization angle random and untraceable. In this thick regime, the emission is depolarized, and the plasma properties can only be constrained to a lower limit.

Measurements of the RM have been used to measure $\dot{M}$ from low-luminosity SMBHs. For example,  \cite{2006ApJ...640..308M,2007ApJ...654L..57M}, hereafter M06 and M07, used observations of Sgr A* with the Submillimeter Array (SMA) to estimate RM,
resulting in the most robust RM estimation to date for Sgr A*, $\mathrm{RM} = (-5.6 \pm 0.7) \times 10^5$ rad m$^{-2}$ on average. Assuming an accretion flow model, they converted the RM into a mass accretion rate, placing a limit of $\leq 2 \times 10^{-7} \, M_{\odot} \, \mathrm{yr}^{-1}$ in the best case scenario, where the magnetic field is near equipartition, ordered and radial.

Similar calculations have been done for M87 following M06/07. For example, \cite{2014ApJ...783L..33K} used SMA observations around 230 GHz, and placed an upper limit $|\mathrm{RM}| < 7 \times 10^{5}$ rad m$^{-2}$, corresponding to  $\dot{M}<9.2 \times 10^{-4} \, M_{\odot} \, \mathrm{yr}^{-1}$ at $21 \, r_{\mathrm{S}}$. In their interpretation, the forward-jet base is the source of polarized emission, while the surrounding spherical flow acts as the Faraday screen. Other works have reported RM measurements in M87 but further along the forward-jet \citep{2002ApJ...566L...9Z,2019ApJ...871..257P}, while others have measured RM from observations at longer wavelengths \citep[7 mm and 1.3 cm, probing regions closer to the jet core,][]{2020A&A...637L...6K}, but the current observing capabilities do not permit sampling sufficiently close wavelengths. 

Using the RM to infer the properties of accretion flows relies on several assumptions. Adopting a spherical flow as the Faraday screen in M87 may be problematic, since we are viewing the source at low inclination. The RM value at 230 GHz reported in \cite{2014ApJ...783L..33K} implies that the EVPA has rotated more than once, breaking the assumptions needed for Equation \ref{eq:rm} to hold. At 230 GHz, it is unclear whether the emission comes from the accretion flow or the jet base, thus it is unclear which one acts as the Faraday screen or as the polarized emission source. It is clear, though, that we can observe linear polarization from the counter-jet at 43 GHz \citep{2018ApJ...855..128W,2021ApJ...922..180P}.

In this work we propose that observations of linear polarization in resolved images of the counter-jet should reveal the presence of the accretion flow onto the black hole. We make predictions for the counter-jet linear polarization and RM from two different methods. 
We first set up an analytic model that uses a cylindrical geometry rather than a spherical geometry to accommodate our view of M87. After predicting the RM from this scenario using Equation \ref{eq:rm}, we examine what the RM value implies about the accretion flow properties (Section \ref{sec:analytic}). In our second method, we employ a polarized radiative transfer code to measure RM from the slope of EVPA vs $\lambda^2$ (Equation \ref{eq:rmobs1}) resulting from two numerical models: a semi-analytic RIAF and a GRMHD snapshot (Section \ref{sec:numerical_models}). This approach allows us control the fundamental physical quantities involved, such as $n_e$, the magnetic field geometry, and the observer's position. We finalize discussing the main findings of each method and their limitations in Section \ref{sec:discussion}.

\section{Analytic Expectations} \label{sec:analytic}

\begin{figure}[t!]
    \centering
    \includegraphics[width=\columnwidth]{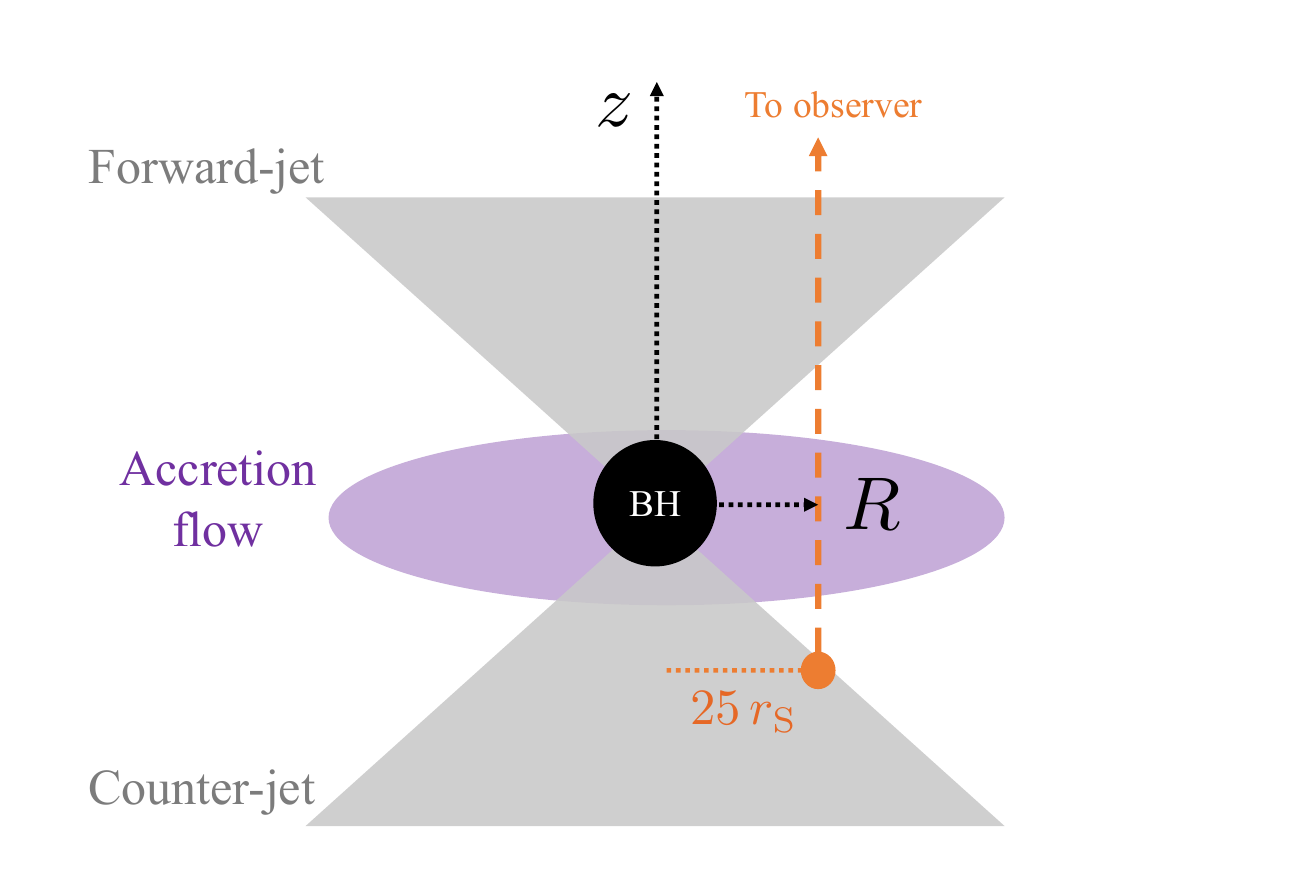}
    \caption{Illustration of our analytical model setup, where our line of sight defines a cylinder of radius $R$ and height $z$. We consider emission originating from a small portion of the counter jet, located at 25 $r_{\mathrm{S}}$ from the black hole. Note that relativistic electrons suppress the forward-jet contribution to the rotation measure, thus we ignore it in our model.}
    \label{fig:analytical_setup}
\end{figure}

We first consider an analytic model of external Faraday rotation of counter-jet emission by the hot accretion flow (Figure \ref{fig:analytical_setup}). Instead of assuming a spherical accretion flow (as previously done by M06/07), we consider a disk in which the density decays with a characteristic height $H$, i.e., the scale height\footnote{The scale height is $H = \Big(\frac{k_B T}{\mu m_H}\Big)^{1/2} \Big( \frac{GM}{R^3}\Big)^{-1/2}$}. The disk is also geometrically thick ($H \sim R$). We then consider cylindrical geometry, such that $n_e$ depends on the cylindrical radius $R$, and height $z$ above the disk mid plane as: 

\begin{equation}
    n_e(R,z) = n_{\mathrm{mid}}(R) e^{-\frac{z^2}{2H^2}},
\end{equation}

\noindent where $n_{\mathrm{mid}}$ is the density at the disk mid plane. To find $n_{\mathrm{mid}}$, we focus on the density of the disk at the mid plane as a function of radius:

\begin{equation}
    \label{eq:densityr}
    n_{\mathrm{mid}}(R) = n_{*} \frac{R_{*}}{R},
\end{equation}

\noindent where $n_{*}$ and $R_{*}$ are normalization values and $R$ is measured in $r_{\mathrm{S}}$. We assume that the disk is embedded in a large scale magnetic field, constant in the $z$-direction, such that there is only a $B_{||}$ component. The $B_{||}$ value depends on the position in the disk at which it is measured. To determine the radial dependence of $B_{||}$, we can compare the magnetic pressure to the gas pressure, related by the plasma beta parameter $\beta = \frac{p_{\mathrm{gas}}}{p_{\mathrm{mag}}}$, which we assume to be constant: 

\begin{equation}
    \beta \frac{B_{||}^2}{8\pi} = \frac{GM\rho}{R},
\end{equation}

\noindent where $G$ is the gravitational constant, and $M$ and $\rho$ are the disk's mass and mass density, respectively. From here one can show that:

\begin{equation}
    \label{eq:bfield}
    B_{||} = 0.14 \, \beta^{-1/2} \Big(\frac{n_e}{\mathrm{cm}^{-3}}\Big)^{1/2} \Big( \frac{R}{r_{\mathrm{S}}} \Big)^{-1/2}.
\end{equation}

\noindent Substituting these equations into Equation \ref{eq:rm}:

\begin{equation}
    \mathrm{RM} = 1.13 \times 10^5 \, \beta^{-1/2} (n_*R_*)^{3/2} R^{-2} \int_{-\infty}^{+\infty} e^{-\frac{3z^2}{4H^2}} dz.
    \label{eq:RMbeforeint}
\end{equation}

\noindent Thus, solving Equation \ref{eq:RMbeforeint} gives the RM as a function of cylindrical radius $R$:

\begin{equation}
    \label{eq:RM_analytical}
    \Bigg( \frac{\mathrm{RM}}{\mathrm{rad \, \, m}^{-2}} \Bigg) \approx 2.27 \times 10^{5} \frac{1}{\beta^{1/2} } \Big( \frac{H}{\mathrm{pc}} \Big)
    \Big( \frac{n_*}{\mathrm{cm}^{-3}} \Big)^{3/2}
\end{equation}
\begin{equation*}
    \Big( \frac{R_*}{r_{\mathrm{S}}} \Big)^{3/2} \Big( \frac{R}{r_{\mathrm{S}}} \Big)^{-2}.
\end{equation*}

In characterizing the cold plasma, we consider a typical density $n_{*} = 3 \times 10^4$ cm$^{-3}$ and $\beta = 10$. Substituting these values into Equation \ref{eq:RM_analytical}, we find that $\mathrm{|RM|} \approx 9.93 \times 10^6$ rad m$^{-2}$, implying $\tau_{\rho_\mathrm{V}} \approx 973$ at 43 GHz. The prediction of $\tau_{\rho_V} \gg 1$ violates the usual assumption of a Faraday thin medium for measuring the RM, and is likely to result in depolarization due to fluctuations in the screen \citep[e.g.,][]{1966MNRAS.133...67B}. In the next section, we carry out full radiative transfer calculations to account for these effects.

\section{Numerical Models} \label{sec:numerical_models}

Next, we use the general relativistic ray tracing code \texttt{grtrans}\footnote{https://github.com/jadexter/grtrans} \citep{2009ApJ...696.1616D,2016MNRAS.462..115D} to solve the polarized radiative transfer equation along ray trajectories for two accretion flow models. The code follows photon trajectories that start from the BH and end at the position of a distant observer's camera. The polarized radiative transfer equation is solved independently along each ray in a curved spacetime, including emission, absorption, and Faraday rotation and its conversion coefficients. The coefficients correspond to those of synchrotron emission from two astrophysical plasmas (or fluid models): (1) a spherical blob, used to mimic point source emission from the counter-jet; and (2) an accretion flow model to mimic that around M87. We first consider a semi-analytical radiatively inefficient accretion flow model, followed by a GRMHD snapshot of the \texttt{koral3D} model.
These three models are described in what follows.

\subsection{Modeling the counter-jet} \label{sec:blob}

The blob model consists of a sphere whose location and size are inputs by the user, as well as the blob's electron density $n_{e,\mathrm{blob}}$, electron temperature $T_{e,\mathrm{blob}}$ and magnetic field strength $B$. For simplicity, we assume that $n_{e,\mathrm{blob}}$, $T_{e,\mathrm{blob}}$ and $B$ are constant inside the blob, and that the magnetic field geometry is entirely radial (so that $B$ is constant along the radial direction). Since the counter-jet emits synchrotron radiation from relativistic electrons, we choose $n_{e,\mathrm{blob}}$ to consist of purely non-thermal electrons ($n_{e,\mathrm{blob}}=n_{e,\mathrm{nth}}$) with a power-law distribution of energies. In this case, the electron energy is no longer described by $T_{e,\mathrm{blob}}$, but instead by the Lorentz factor $\gamma$, which we set to 100 to ensure relativistic electrons in the counter-jet. Thus, the blob's emissivities are calculated for a polarized synchrotron power-law source.

\begin{figure}[t!]
    \centering
    \includegraphics[width=\columnwidth]{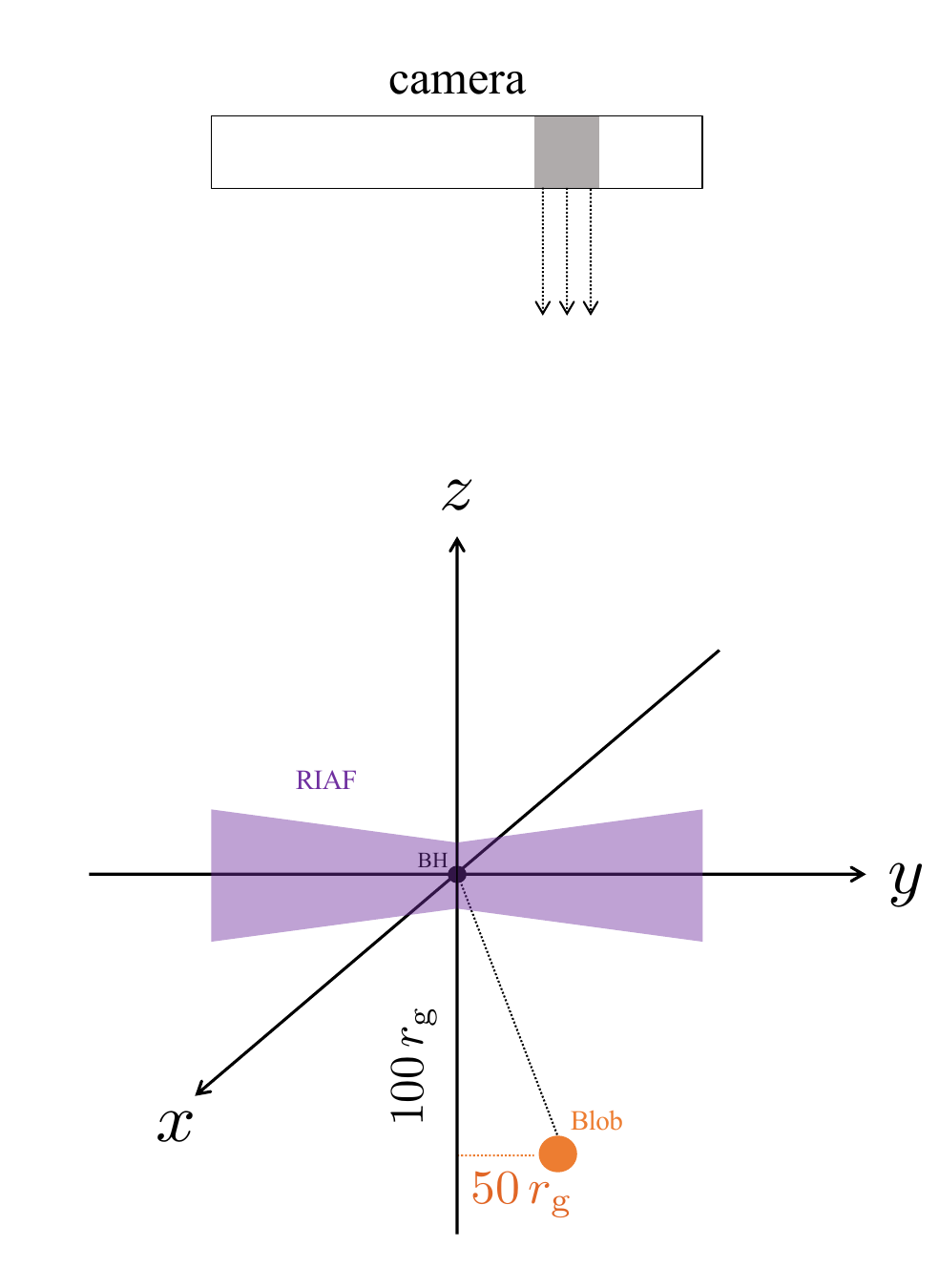}
    \caption{We consider a face-on observer looking straight down into a geometrically thick accretion flow surrounding the black hole. The accretion flow electrons are assumed to be non-relativistic, while the observed synchrotron emission results from a radiating blob. The shaded area indicates where the blob will appear on the camera, with intensities calculated by backwards ray tracing. The blob's radius is 2 $r_{\mathrm{g}}$, and it is placed at 100 $r_{\mathrm{g}}$ in vertical distance and 50 $r_{\mathrm{g}}$ in horizontal distance (equivalent to 25 $r_{\mathrm{S}}$) from the black hole to mimic our analytical model. For \texttt{koral3D}, we use the same emission setup and replace only the accretion flow model.}
    \label{fig:numerical_setup}
\end{figure}

The blob's center is located at $r_{\mathrm{blob}}=100 \, r_\mathrm{g}$, $\theta_{\mathrm{blob}}=2.62$ and $\phi_{\mathrm{blob}}=0$, such that the blob is located at $50 \, r_{\mathrm{g}}$ (or equivalently, $25 \, r_{\mathrm{S}}$, see Figure \ref{fig:numerical_setup}) as seen from an observer above. The blob's properties are set as follows: $n_{e,\mathrm{blob}}=10^4$ cm$^{-3}$, and $B_{\mathrm{blob}}=0.1$ G. Because we want to sample a distinct cylindrical radius $R$, the size of the blob is chosen to be small (2 $r_{\mathrm{g}}$)\footnote{The gravitational radius is related to the Schwarzschild radius by $1 \, r_{\mathrm{S}}= 2 r_{\mathrm{g}}$.}. The blob is placed at an angle of 150$^{\circ}$ measured from the positive $\hat{z}-$axis, and at a distance of 100 $r_{\mathrm{g}}$ from the BH along $\hat{z}$, such that the blob is located at $R=25 \, r_{\mathrm{S}}$ from the BH axis as seen from the observer above. We use a small camera centered on the blob in order to concentrate resolution there and ensure converged results. The simulation is conducted for a BH mass of $6.5 \times 10^5 \, M_{\odot}$. 

\subsection{RIAF} \label{sec:riaf}

\begin{figure*}[t]
    \centering
    \includegraphics[width=\textwidth]{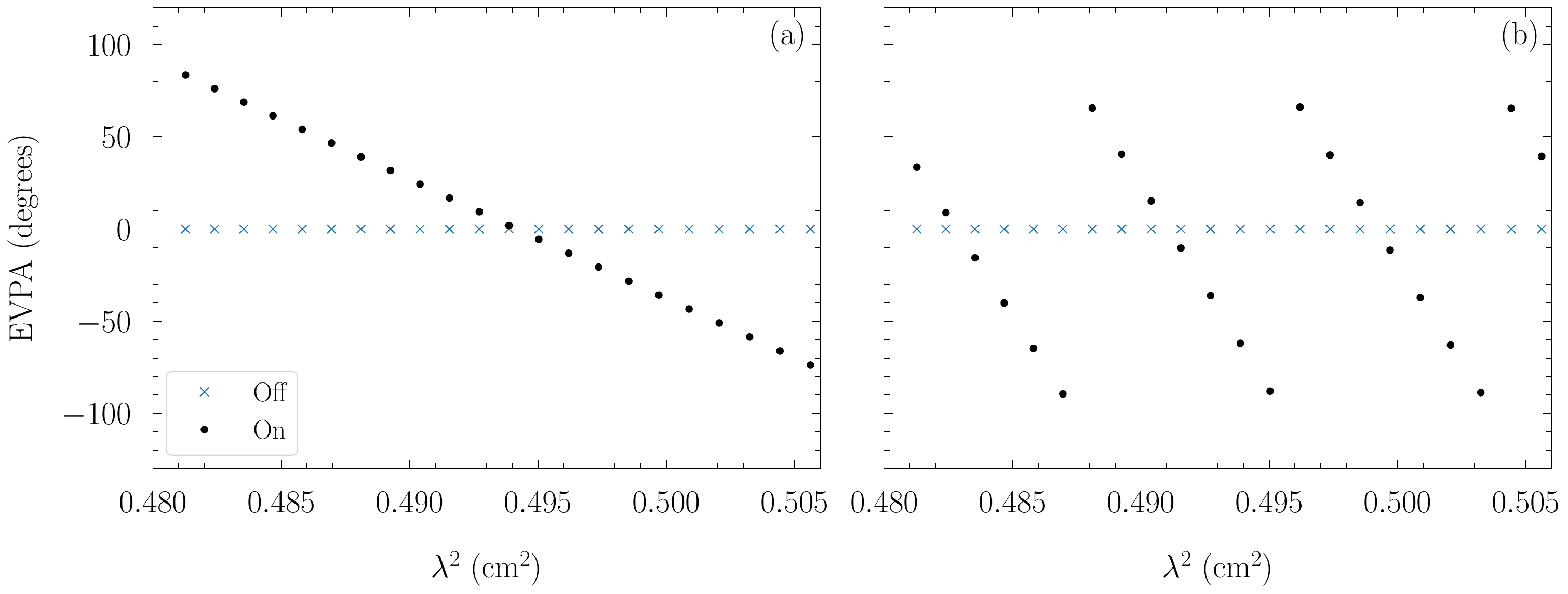}
    \caption{Polarization position angle measured at different frequencies, spanning $\sim 4.2\times 10^{10}-4.3\times 10^{10}$ Hz, when the Faraday rotation effect is on (black filled circles) and off (blue crosses). \textit{Panel a} displays measurements from a simulation ran using the RIAF model, while \textit{panel b} corresponds to the \texttt{koral3D} model results. In both cases, the position of the blob is $\phi_{\mathrm{blob}}=0$. When the Faraday rotation effect is off, the polarization angle remains unchanged.}
    \label{fig:faraday_off}
\end{figure*}

\subsubsection{Model description and simulation setup}

The semi-analytical radiatively inefficient accretion flow \texttt{sariaf} describes a hot, quasi-spherical rotating flow in which a small fraction of the gravitational energy is radiated away \citep[][hereafter B09]{2006ApJ...636L.109B,2009ApJ...697...45B}. The emission mechanism is synchrotron and can originate from both thermal and non-thermal electrons, although in our model, we choose thermal electrons only ($n_e=n_{e,\mathrm{th}}$). In a RIAF, the accretion flow is geometrically thick, $H\sim R$, and characterized $n_{e,\mathrm{th}}$, $T_e$ and $B$. As described in B09, these are given by:

\begin{equation}
    \label{eq:n_b09}
    n_{e,\mathrm{th}} = n^0_{e,\mathrm{th}} \Big( \frac{R}{R_\mathrm{S}} \Big)^{-1.1} e^{-z^2/2H^2}
\end{equation}

\begin{equation}
    \label{eq:t_b09}
    T_e = T^0_e \Big( \frac{R}{R_\mathrm{S}} \Big)^{-0.84}
\end{equation}

\begin{equation}
    \label{eq:bfield_b09}
    \frac{B^2}{8\pi} = \beta^{-1} n_{e,\mathrm{th}} \frac{m_p c^2 r_{\mathrm{S}}}{12 R}
\end{equation}

\noindent where $m_p$ is the proton mass and the superscript 0 denotes normalization values. Although in the \texttt{sariaf} model the magnetic field geometry is toroidal ($\phi$-direction), we accommodate it to have purely poloidal ($\theta$-direction) geometry instead. In that way, we ensure that the magnetic field has a component directed along the line of sight (in the vertical direction) which is needed for Faraday rotation to occur.

To mimic our analytical set up, we employ the \texttt{blob} model to represent point source-like emission from the counter-jet and a RIAF model to represent the accretion flow in M87. We do not consider the forward-jet, since relativistic electrons suppress the forward-jet contribution to the RM \citep{2000ApJ...545..842Q}. For this purpose, we combine the \texttt{blob} and \texttt{sariaf} models into a new fluid model in \texttt{grtrans}, called \texttt{sariafblob}. As shown in Figure \ref{fig:numerical_setup}, the camera is located face-on with respect to the BH and accretion flow axis, such that the observer is looking straight down, as described in Section \ref{sec:blob}.

The inputs for the \texttt{sariafblob} model correspond to the normalization values $n_{e,\mathrm{th}}^0$, $T_{e}^0$ at $1 \, r_{\mathrm{S}}$, and $\beta$, which allows us to determine the magnetic field value. Then, density, temperature and magnetic field vary as a function of radius according to Equations \ref{eq:n_b09}, \ref{eq:t_b09}, and \ref{eq:bfield_b09}, respectively. These input parameters depend on our choice of normalizations and, unlike the analytic formula, the code requires an electron temperature normalization $T_e^0$.

Since the accretion flow of M87 is not detected at radio wavelengths, we assume that the accretion flow electrons are non-relativistic and set $T_{e}^0 = 5 \times 10^9$ K. We further choose $n_{e,\mathrm{th}}^0=3\times 10^4$ cm$^{-3}$ and $\beta=10$ (the default), consistent with our choice in the analytic formula. 

\subsubsection{Model results} \label{subsubsec:riaf_results}

We first confirm that Faraday rotation is the only physical mechanism causing the rotation of the EVPA. Thus, if we turn off the Faraday rotation effect, we should expect no rotation in the EVPA. This is confirmed in Figure \ref{fig:faraday_off}, \textit{panel a}, represented by blue crosses.

The EVPA rotation as a function of wavelength squared for the RIAF model is also shown in Figure \ref{fig:faraday_off}, \textit{panel a}, marked by black filled circles. We fit a line to these data points and obtain $|\mathrm{RM}| \approx 1.13 \times 10^6$ rad m$^{-2}$. This figure distinctly shows the linear dependence $\mathrm{EVPA}\sim\lambda^2$, supporting the external Faraday rotation scenario in which a uniform Faraday screen rotates the EVPA. The RM value implies that $\tau_{\rho_\mathrm{V}} \approx 111$ at 43 GHz, matching the code's calculation. Compared to the analytical result, these values are lower but confirm the prediction of a Faraday thick accretion flow. We also calculate an average degree of linear polarization of $\sim 9\%$ over the wavelength range, implying that the blob's emission is depolarized, again supporting the analytical prediction.

Additionally, we perform two simple tests to reaffirm that the accretion flow acts as the Faraday screen, with the blob not contributing to the RM. Since $\mathrm{RM} \propto n_e^{3/2}$, decreasing the RIAF density by a factor 100 should decrease the RM by factor 1000. To conduct this test, we compare the best-fit slopes for a run with $n_{e,\mathrm{th}}^0=3\times 10^4$ cm$^{-3}$ and $n_{e,\mathrm{th}}^0=3\times 10^2$ cm$^{-3}$. The former corresponds to $|\mathrm{RM}| \approx 1.13 \times 10^6$ rad m$^{-2}$ as mentioned earlier, while the latter yields $|\mathrm{RM}| = 1.67 \times 10^3$ rad m$^{-2}$. The factor 1000 decrease in RM implies that the accretion flow \textit{is} the Faraday screen, confirming the external Faraday rotation scenario in which the rotation of the plane of polarization occurs \textit{outside} the emission source.
Similarly, if the blob is not significantly contributing to the RM, reducing its density should have a minimal impact on the previous result. For a blob with $n_{e,\mathrm{blob}}=10^2$ cm$^{-3}$, $|\mathrm{RM}| = 1.13 \times 10^6$ rad m$^{-2}$, indicating that the RM remains unchanged, further supporting our conclusion.

While we considered the face-on scenario to compare with our analytical model, M87 is at an inclination of $17^{\circ}$. Therefore, it is worth examining the effects of varying the camera inclination on the EVPA and linear polarization degree. These results are shown in Figure \ref{fig:inclined_blob}, Section \ref{sec:inclination}. An inclined line of sight does not alter the EVPA linearity or the conclusion that the accretion flow is in the Faraday thick regime. However, with a purely vertical magnetic field, depolarization is significantly lower than in the face-on case. This result highlights the need for more realistic field geometries, which we explore in the following section.

\subsection{koral} \label{sec:koral}

\begin{figure*}
    \centering
    \includegraphics[width=\textwidth]{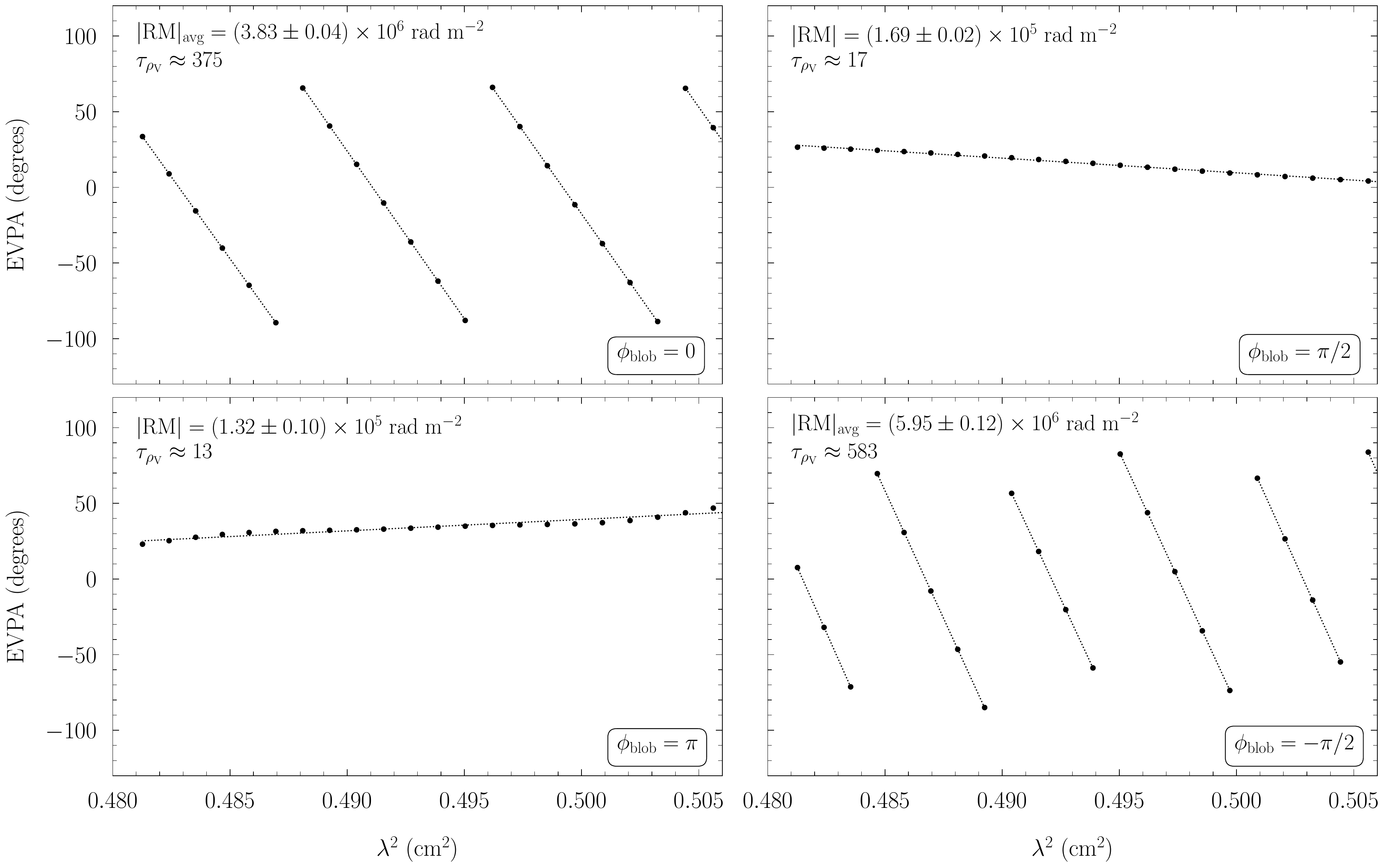}
    \caption{Polarization position angle measured at different frequencies, spanning the same range as in Figure \ref{fig:faraday_off}, obtained from the \texttt{koral3D} model. The panels displays EVPA as a function of $\lambda^2$ for each blob position in azimuthal angle, as indicated in the lower right corner. The dotted lines corresponds to the best-fit RM, and we show its values in the upper left corner together with the corresponding $\tau_{\rho_\mathrm{V}}$. When multiple stripes are observed, we quote the average RM instead (where the error is the standard deviation) and the $\tau_{\rho_\mathrm{V}}$ reported is at 43 GHz. While in all cases we observe RM signatures, these vary according to the blob position.}
    \label{fig:koral_evpa}
\end{figure*}

\begin{figure*}
    \centering
    \includegraphics[width=\columnwidth]{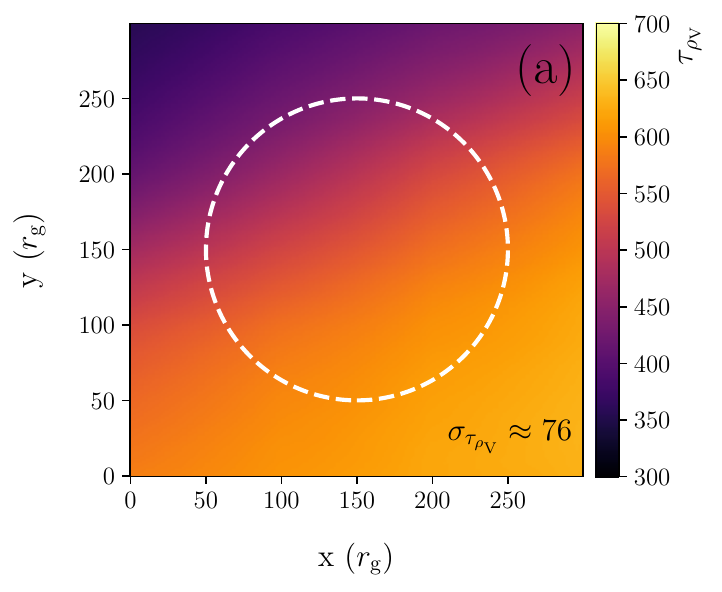}
    \includegraphics[width=\columnwidth]{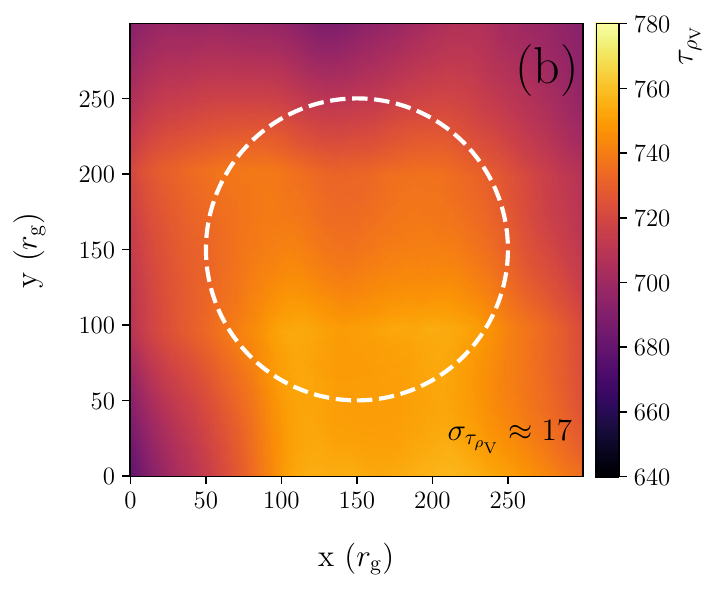}
    \includegraphics[width=\columnwidth]{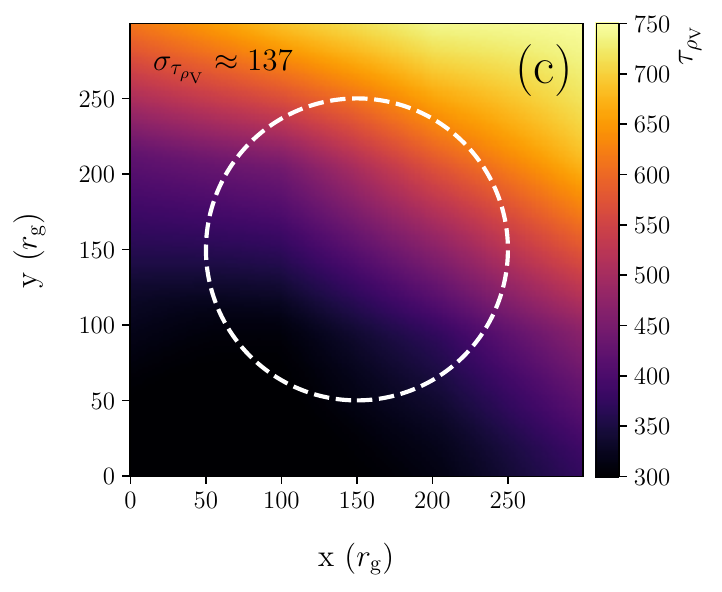}
    \includegraphics[width=\columnwidth]{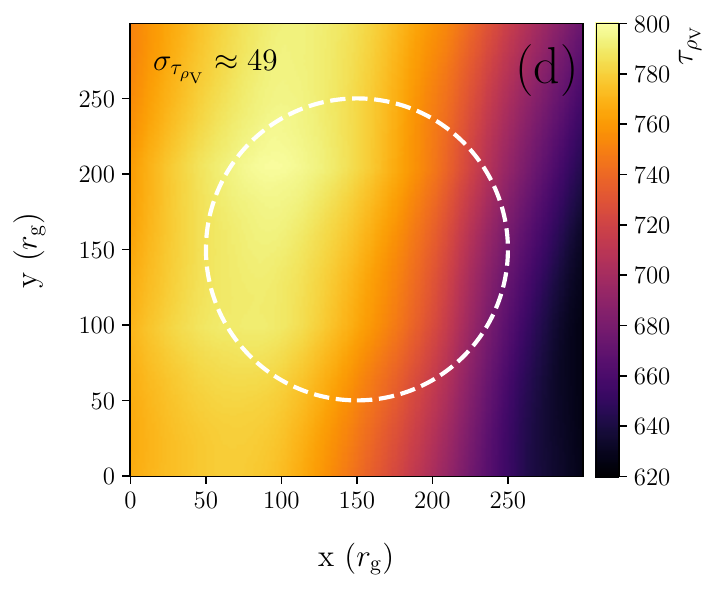}
    \caption{Maps of $\tau_{\rho_\mathrm{V}}$ at 43 GHz for each position in azimuthal angle. \textit{Panel a} is $\phi_{\mathrm{blob}}=0$, \textit{panel b} is $\phi_{\mathrm{blob}}=\pi/2$, \textit{panel c} is $\phi_{\mathrm{blob}}=\pi$ and \textit{panel d} is $\phi_{\mathrm{blob}}=-\pi/2$, as in in Figure \ref{fig:koral_evpa}. The white dashed line represents the approximate region occupied by the blob. The color bar displays Faraday depth values, where limits are different in each panel for better visualization of the $\tau_{\rho_\mathrm{V}}$ gradients. We include the $\sigma_{\tau_{\rho_\mathrm{V}}}$ values, which represent the fluctuation of $\tau_{\rho_\mathrm{V}}$ around its mean value over the region showed. Each position, corresponding to a different line of sight, experiences different Faraday depth gradients and fluctuations across the blob region. In all cases, the accretion flow is Faraday thick, with $\tau_{\rho_\mathrm{V}} \gg 1$.}
    \label{fig:taurhoV_map}
\end{figure*}

\begin{figure}
    \centering
    \includegraphics[width=\columnwidth]{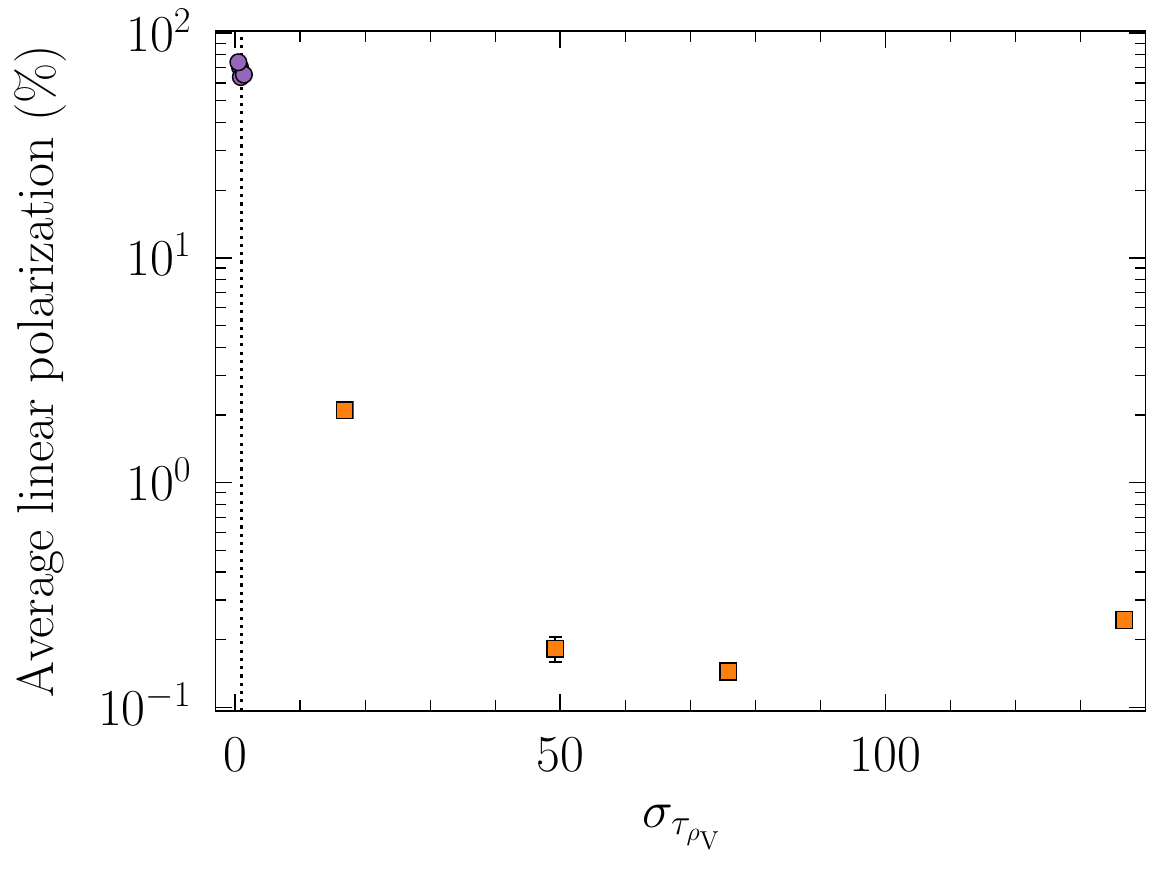}
    \caption{Linear polarization degree as a function of Faraday depth fluctuation $\sigma_{\tau_{\rho_\mathrm{V}}}$. The orange squares correspond to the blob positions displayed in Figures \ref{fig:koral_evpa} and \ref{fig:taurhoV_map}, where we averaged the degree of polarization over the range of frequencies considered. The purple points correspond to the same measurements when imposing $\sigma_{\tau_{\rho_\mathrm{V}}}=1$, marked by the vertical dotted line. Note that, in all cases, the accretion flow is Faraday thick, with $\tau_{\rho_\mathrm{V}} \gg 1$. The decay in linear polarization degree with increasing fluctuations, together with the polarization recovery in the Faraday thin limit, suggest that $\sigma_{\tau_{\rho_\mathrm{V}}}$ plays a major role in depolarizing the emission.}
    \label{fig:pol_degree}
\end{figure}

\subsubsection{Model description}

As a final example of possible M87 accretion flow properties, we consider a single late time snapshot of a radiation general relativistic MHD simulation of M87 \citep{chael2019} run using the code \texttt{koral3D} \citep{sadowski2017}. The simulation solves the equations of ideal MHD in a fixed Kerr spacetime assuming an initial gas torus in hydrodynamic equilibrium, with electron heating and radiative cooling for system paramaters varied to produce a spectral energy distribution (luminosity) comparable to that of M87. We take from the simulation snapshot the magnetic field strength and geometry, particle density, and gas pressure all in cgs units. Since the calculation is 3D and fully time-dependent, it includes fluctuations due to turbulence driven by the magnetorotational instability \citep{mri} which are not present in the RIAF model considered above.

\subsubsection{Model results}

Following the procedure described in Section \ref{subsubsec:riaf_results}, we first confirm that Faraday rotation is the only effect responsible for rotating the EVPA. This is shown in Figure \ref{fig:faraday_off}. Because \texttt{koral3D} is inherently different in nature than the ordered RIAF (e.g., magnetic field geometry), we can expect a different EVPA behavior depending on the blob position. Thus, we focus on four azimuthal positions: $\phi_{\mathrm{blob}}=0,\pi/2,\pi,-\pi/2$, where $\phi_{\mathrm{blob}}=0$ is the position we assumed in the RIAF model. The EVPA as a function of $\lambda^2$ for different blob positions is shown in Figure \ref{fig:koral_evpa}. For two azimuthal angles ($\phi_{\mathrm{blob}}=\pi/2,\pi$) we observe a single stripe, as in the RIAF case, but with flatter slopes. The other two positions ($\phi_{\mathrm{blob}}=0,-\pi/2$) display multiple stripes, but with RMs closer in value to that found in RIAF. In general we observe a linear dependence in all cases, consistent with external Faraday rotation, and allowing us to use $\tau_{\rho_{\mathrm{V}}} = 2 \, \mathrm{RM} \lambda^2$. The RMs and corresponding $\tau_{\rho_{\mathrm{V}}}$ are included in each panel for each of the blob positions. In all cases $\tau_{\rho_{\mathrm{V}}}\gg 1$, suggesting a Faraday thick accretion flow.

However, the RM value (slope) and therefore the inferred $\tau_{\rho_\mathrm{V}}$ vary with blob position. To understand this variable behavior, we explore the values of $\tau_{\rho_\mathrm{V}}$ near the blob region, shown in Figure \ref{fig:taurhoV_map}. We find that $\tau_{\rho_\mathrm{V}}$ varies with blob position, and that there is also a $\tau_{\rho_\mathrm{V}}$ gradient within the blob region (marked with dashed white line). The approximate value of such gradient is reported as $\sigma_{\tau_{\rho_\mathrm{V}}}$ in each panel. For our particular \texttt{koral3D} snapshot, $\sigma_{\tau_{\rho_\mathrm{V}}}/\tau_{\rho_{\mathrm{V}}}$ ranges from 0.1 to 11, suggesting that the size of the fluctuations play a role when $\tau_{\rho_\mathrm{V}} \gg 1$, as it is the case in all of our simulations. Thus, each blob position experiences a different Faraday screen, resulting in different EVPA behavior and RM value. The fact that each line of sight ``observes" different Faraday depths can result in depolarization and scramble the EVPA \citep{1966MNRAS.133...67B}. Figure \ref{fig:pol_degree} shows the degree of linear polarization for each blob position (orange squares), showing indeed the low level of polarization (or high depolarization). However, as shown by Figure \ref{fig:koral_evpa}, depolarization does not scramble the EVPA pattern, in turn allowing us to still observe the linear behavior.

In trying to understand which parameter determines the depolarization, i.e., whether it is $\tau_{\rho_\mathrm{V}}$ or  $\sigma_{\tau_{\rho_\mathrm{V}}}$, we conduct another test in which we impose the Faraday thin limit by decreasing the accretion flow density by $\sim 100$. We then examine the $\tau_{\rho_\mathrm{V}}$ range on each map, the $\sigma_{\tau_{\rho_\mathrm{V}}}$ within the map, and polarization degree. The results of these tests are shown in Figure \ref{fig:pol_degree} with purple circles. In all cases, we find that the linear polarization degree is recovered in the thin limit, well above $\sim 60\%$, while $\tau_{\rho_\mathrm{V}} \gg 1$. In other words, we observe that a Faraday thin accretion flow becomes polarized again so long as $\sigma_{\tau_{\rho_\mathrm{V}}}\sim 1$.
\newpage

\section{Discussion} \label{sec:discussion}

\subsection{Summary of results}

We propose that resolved radio images of linearly polarized emission from the M87 counter-jet \citep[e.g., similar to][]{2021ApJ...922..180P} may reveal the presence of the accretion flow through the Faraday rotation effect. Such measurements have the potential to constrain the accretion flow properties as quantified by the Faraday rotation measure (RM) or depolarization of the counter-jet emission.

From an analytic calculation (Section \ref{sec:analytic}) we find that at 43 GHz, $\mathrm{|RM|} \approx 9.93 \times 10^6$ rad m$^{-2}$, implying $\tau_{\rho_\mathrm{V}} \approx 973$. These results imply a Faraday thick accretion flow, where the polarization angle cannot be traced back to its origin, leading to complete depolarization of the counter-jet emission and possibly no visible RM signatures. To test this prediction and explore more realistic scenarios, we used the general relativistic ray tracing code \texttt{grtrans}.

In the numerical models, we considered a blob to mimic the counter-jet emission, allowing us to study the impact of including multiple lines of sight in our calculations. The accretion flow was modeled either as a RIAF or with \texttt{koral3D}, and we analyzed the polarization angle’s behavior as a function of $\lambda^2$ in each case.
The RIAF models show the $\mathrm{EVPA} \sim \lambda^2$ trend expected for external Faraday rotation, with the slope corresponding to the RM. At 43 GHz, we find $|\mathrm{RM}| \approx 1.13 \times 10^6$ rad m$^{-2}$ and $\tau_{\rho_\mathrm{V}} \approx 111$, consistent with the code's calculation. Although these values are lower than the analytical estimates, they confirm the Faraday-thick regime. In this regime we should expect the EVPA to result in a completely random position, but instead we often observe that the linearity persists. The average linear polarization degree across the wavelength range is $\sim 9\%$, confirming that the blob's emission is partially depolarized.

The \texttt{koral3D} model is more realistic as it includes non-zero turbulence and fluctuations. In this case, we find that the EVPA behavior changes with the blob position in azimuthal angle. Remarkably, in each position, we still observe $\mathrm{EVPA} \sim \lambda^2$, the linear pattern consistent with external Faraday rotation, and that allows us to fit the RM. In general, the inferred RMs and $\tau_{\rho_\mathrm{V}}$ are lower than the analytic predictions, and the $\tau_{\rho_\mathrm{V}}$ values from the RM fit do not match the code calculations shown in the maps. To understand the azimuthal variability, we examine the $\tau_{\rho_\mathrm{V}}$ maps in and around the blob region. We find that the maps are different among the blob positions, and also each individual map displays a $\tau_{\rho_\mathrm{V}}$ gradient within the blob region. This result indicates that different lines of sight experience varying Faraday depths, with fluctuation scales in $\tau_{\rho_\mathrm{V}}$ also varying. Although the accretion flow remains Faraday thick with significant $\tau_{\rho_\mathrm{V}}$ fluctuations, we still observe linearity in EVPA. We also find that the degree of linear polarization is low ($\lesssim3\%$) in all cases, confirming the blob emission is depolarized, as predicted by the analytic model.

\subsection{Implications for M87} \label{subsec:implications}

All models considered predict that the accretion flow should have an observable impact on the linear polarization from the counter-jet, by rotating the EVPA and/or depolarizing the counter-jet emission. Observations of spatially resolved polarization could therefore confirm the presence of an accretion flow. However, since we expect the accretion flow to be Faraday thick, determining its properties ($n_e$, $B_{||}$, $\dot{M}$) is challenging because the EVPA cannot be traced back to its original position, unlike in a Faraday-thin scenario. Previous studies, including those of the Galactic center and M87 \citep[M06/07,][]{2014ApJ...783L..33K}, often assume the accretion flow is Faraday thin, using polarization data to fit RM and estimate $\dot{M}$ based on the linearity of $\mathrm{EVPA} \sim \lambda^2$. This assumption implies that the EVPA has rotated by less than $180^\circ$, allowing use of Equation 
\ref{eq:rm}. However, we have shown that a Faraday thick accretion flow can still display a linear behavior. In M87, this misinterpretation could lead to inaccurate conclusions about the accretion flow properties, while in reality, the observations may correspond to a $\tau_{\rho_\mathrm{V}} = 1$ surface.

Focusing on our more realistic \texttt{koral3D} test, the results suggest significant variability in the Faraday screen, as demonstrated by the changing $\tau_{\rho_\mathrm{V}}$ gradients with different blob position, even within a single GRMHD snapshot. Moreover, $\tau_{\rho_\mathrm{V}}$ can fluctuate across the region of interest, which we quantify with $\sigma_{\tau_{\rho_\mathrm{V}}}$. Considering the variability of $\tau_{\rho_\mathrm{V}}$, observed EVPAs must be interpreted carefully, since this variability can hinder the underlying RM and true polarization state. 

Our simulations emphasize the complexity of Faraday rotation in systems like M87. In the Faraday-thick regime, we observe that the emission can depolarize significantly in some cases and less so in others. We calculate $\sigma_{\tau_{\rho_\mathrm{V}}}$ and find that depolarization becomes large when $\sigma_{\tau_{\rho_\mathrm{V}}}\gg 1$ (see Figure \ref{fig:pol_degree}), i.e, where the EVPA along different lines of sight undergoes a different number of rotations. Interestingly, even with significant depolarization, the linearity of the EVPA with $\lambda^2$ persists, as previously studied by \cite{1966MNRAS.133...67B}. We note that this finding does not imply that the EVPA is always linear; rather, it indicates that linearity can be observed even in cases with high depolarization.

Finally, one direct application of the RM is predicting $\dot{M}$. From our analytical expression for RM in Equation \ref{eq:RM_analytical}, this is done by assuming a density profile and solving the line integral, which requires a Faraday thin accretion flow, as outlined by M06/07 and \cite{2014ApJ...783L..33K}. However, in the Faraday thick case (like in our simulations) $\dot{M}$ cannot be measured accurately. Still, even in this scenario, valuable information can be obtained because (1) it serves as evidence of an accretion flow depolarizing the counter-jet emission, and (2) a RM measurement implies a lower limit on $\tau_{\rho_{\mathrm{V}}}$, density, and in turn, a lower limit on $\dot{M}$. Because RM displays variability in the simulations, the lower limit measurements can be refined and later compared to detailed numerical models to ultimately infer $\dot{M}$. These methods allow us to determine $\dot{M}$ independently of other approaches, for example, using X-ray emission to infer $\dot{M}$ at the Bondi radius, or EHT observations at horizon scales.

\subsection{Limitations and open questions}

\begin{figure}
    \centering
    \includegraphics[width=1.04\columnwidth]{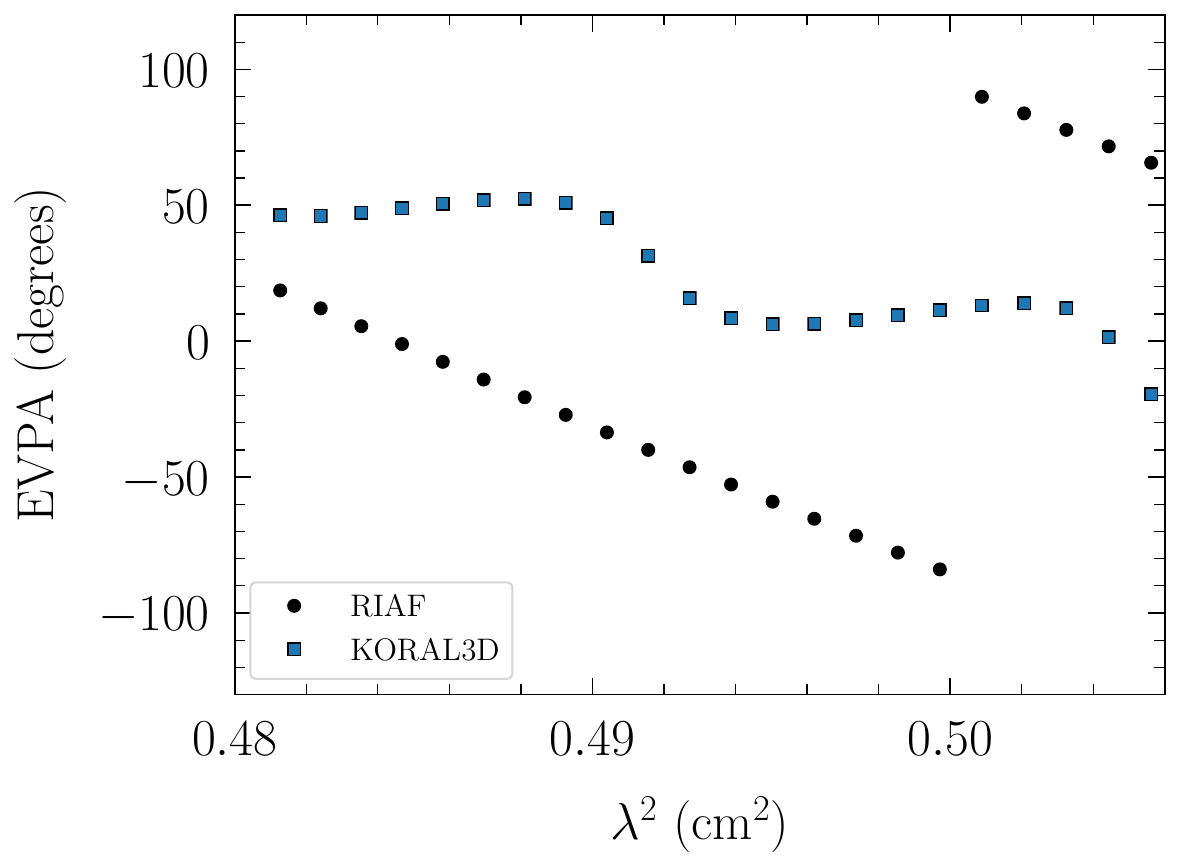}
    \caption{Polarization position angle measured at different frequencies for the RIAF (black circles) and \texttt{koral3D} (blue squares) models with a larger blob size, and spanning the same range as in Figures \ref{fig:faraday_off} and \ref{fig:koral_evpa}. In these examples, the blob is positioned at $\phi_{\mathrm{blob}}=\pi/2$ with a radius of 10 $r_\mathrm{g}$. While the RIAF model retains a linear EVPA, the \texttt{koral3D} model displays deviations due to additional lines of sight and $\tau_{\rho_{\mathrm{V}}}$ fluctuations.}
    \label{fig:big_blob}
\end{figure}

All three models we have tested operate under the assumption that the counter-jet is visible, making them exclusively applicable to M87 at longer wavelengths, when this condition holds true (i.e., not at 230 GHz, where the image appears as a diffuse ring). Furthermore, because there is no real image of the counter-jet, we assume the size of the emission region and its morphology. A larger emission region could amplify the effect of fluctuations in $\tau_{\rho_\mathrm{V}}$, as it would include more lines of sight, each rotating the EVPA by different amounts, which could disrupt the linear $\lambda^2$ pattern. Figure \ref{fig:big_blob} illustrates this scenario for the RIAF and \texttt{koral3D} models. The blob is located at $\phi_{\mathrm{blob}}=\pi/2$ and its radius is increased to 10 $r_\mathrm{g}$. The RIAF model retains a linear EVPA, as expected from the uniformity of this model, while the \texttt{koral3D} model shows clear deviations. In both cases, the degree of linear polarization is $\lesssim 1\%$, confirming that counter-jet emission should be depolarized, as in our previous results. 

Our more realistic test with \texttt{koral3D}, uses a single GRMHD snapshot of M87. This model assumes specific accretion flow properties for M87 at 50 $r_\mathrm{g}$, the region of interest. However, $T_e$, $n_e$ and the magnetic field strength and geometry are not well known at this distance. These values are thus scaled in the simulations. As discussed in Section \ref{subsec:implications}, modifying the scaling of $n_e$ alters the observations, as the linear polarization is recovered for lower $n_e$. By using a single GRMHD snapshot, we did not account for time variability, although this is hinted at by the changing EVPA behavior across different azimuthal angles. This changing behavior suggests that the RM signatures will vary depending on when and where we observe the nuclear region of M87, which in turn impacts our $\dot{M}$ predictions. Conducting these tests with other GRMHD models of M87 and including a time study are beyond the scope of this work, but highly encouraged to confirm our results.

While the primary goal of the GRMHD snapshot is to examine the screen’s imprint on the RM signature, it can also be used to model the counter-jet, incorporating physics absent in the blob model (e.g., variations in counter-jet size at different radio wavelengths). However, making a realistic-looking jet at 43 GHz would introduce additional uncertainties to our controlled blob experiment, which is directly comparable to our analytical model. First, GRMHD models of thermal jet emission tend to underproduce the emission region size, for example, in Sgr A* \citep{issaoun2019}, and therefore a non-thermal electron population is likely needed \citep[e.g.,][]{ozel2000}. Second, GRMHD models including non-thermal electrons are in their early stages \citep[see, e.g.,][]{davelaar2019}. Therefore, the results would likely depend on the details of the assumed spatial and energy distributions of non-thermal particles. While modeling the jet is beyond the scope of this work, it will be essential to revisit these experiments once realistic jet models become available. In the meantime, other ways to explore the impacts of the counter-jet emission region include a more realistic magnetic field configuration within the blob, or placing multiple smaller blobs to mimic the emission along the counter-jet. Assuming identical intrinsic emission properties across all blobs, the latter approach is expected to amplify the effect of fluctuations in $\tau_{\rho_\mathrm{V}}$ (as observed in the 10 $r_\mathrm{g}$ blob test) and disrupt the linearity in $\lambda^2$.

One of our key findings is that the counter-jet emission is depolarized, yet the EVPA can still display a linear dependence with $\lambda^2$. However, this does not imply that EVPA always retains its linear dependence in the Faraday thick regime with high depolarization, only that it \emph{can} under certain conditions. Further studies of the Faraday thick regime are needed to understand when this is the case, and whether depolarization is determined only by $\sigma_{\tau_{\rho_\mathrm{V}}}$ rather than $\tau_{\rho_\mathrm{V}}$, as our results suggest.

Regardless, the M87 accretion flow should be readily visible as depolarization (and possibly detectable Faraday rotation) of the counter-jet in polarized Very Long Baseline Interferometry observations at radio frequencies  \citep[e.g.,][]{2021ApJ...922..180P}. Similarly, in any low-luminosity SMBH where the forward- and counter-jet can be resolved separately in linear polarization, the methods we have used can be applied to study the accretion flow in such systems. One promising example is NGC 1052, where the counter- and forward-jet are resolved in images at 43 GHz \citep{2016A&A...593A..47B,2019A&A...623A..27B}. Future observations may also make measurements or set lower limits on the particle density and mass accretion rate onto the black hole on small scales, close to the event horizon.

This work was supported in part by the National Science Foundation through the awards AST-1909711 and AST-2307983, and through an Alfred P. Sloan Fellowship (JD). 

\vspace{5mm}

\software{\texttt{grtrans}    
          }
\newpage
\appendix
\section{The effect of camera inclination on the RIAF model} \label{sec:inclination}
\begin{figure}[h!]
    \centering
    \includegraphics[width=0.55\textwidth]{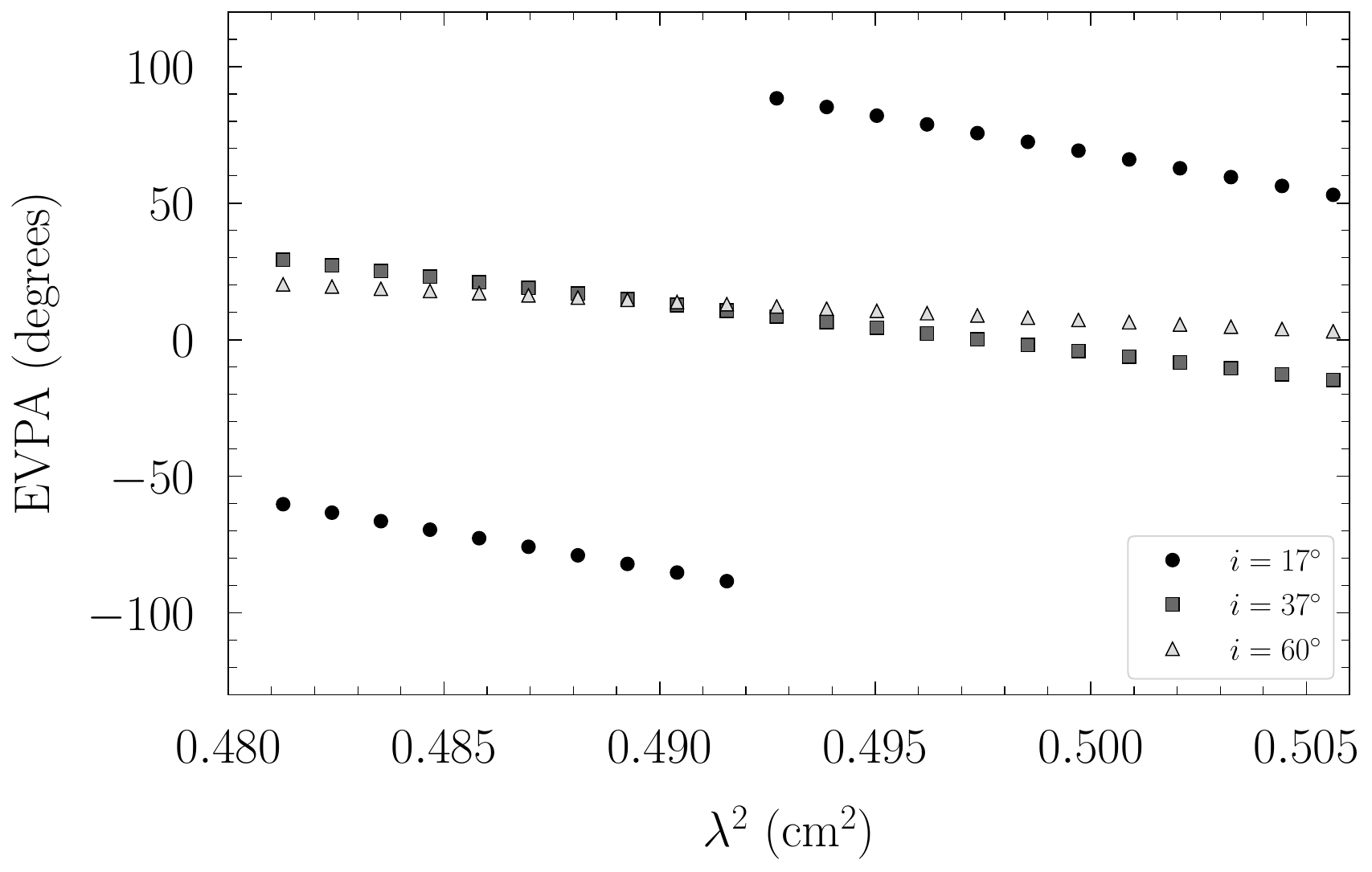}
    \caption{Polarization position angle measured at different frequencies for the RIAF model with different camera inclinations. Using the setup from Section \ref{sec:blob}, the inclination is adjusted to $17^{\circ}$ (black circles), $37^{\circ}$ (dark grey squares), and $60^{\circ}$ (light grey triangles) while maintaining a constant projected distance of 50 $r_\mathrm{g}$. In the RIAF model, the magnetic field is purely in the z-direction, such that higher inclinations reduce its contribution, producing flatter EVPAs and lower RMs ($\mathrm{|RM|} = 1.2-4.8 \times 10^5$ rad m$^{-2}$) while still in the Faraday thick regime ( $\tau_{\rho_V} \gg 1$). In these three examples, the degree of linear polarization increases to $>60\%$, indicating reduced depolarization compared to the face-on case ($9\%$). This result highlights the risk of misinterpreting depolarization when assuming a purely vertical, unrealistic magnetic field, and reinforces the importance of using more realistic field geometries, such as those in our GRMHD snapshot.}
    \label{fig:inclined_blob}
\end{figure}

\newpage
\bibliography{M87_FR}{}
\bibliographystyle{aasjournal}

\end{document}